\documentclass[manuscript]
{acmart}
\usepackage{graphicx} 
\usepackage{multirow}

\begin{document}
\title{Melody: A Platform for Linked Open Data Visualisation and Curated Storytelling}

\author{Giulia Renda}
\email{giulia.renda3@unibo.it}
\orcid{0000-0003-2990-0021}
\author{Marilena Daquino}
\email{marilena.daquino2@unibo.it}
\orcid{0000-0002-1113-7550}
\author{Valentina Presutti}
\email{valentina.presutti@unibo.it}
\orcid{0000-0002-9380-5160}
\affiliation{%
  \institution{University of Bologna}
  \country{Italy}
}

\begin{abstract}
    Data visualisation and storytelling techniques help experts highlight relations between data and share complex information with a broad audience. However, existing solutions targeted to Linked Open Data visualisation have several restrictions and lack the narrative element. In this article we present MELODY, a web interface for authoring data stories based on Linked Open Data. MELODY has been designed using a novel methodology that harmonises existing Ontology Design and User Experience methodologies (eXtreme Design and Design Thinking), and provides reusable User Interface components to create and publish web-ready article-alike documents based on data retrievable from any SPARQL endpoint. We evaluate the software by comparing it with existing solutions, and we show its potential impact in projects where data dissemination is crucial.
\end{abstract}


\maketitle

\section{Introduction}
Nowadays, Linked Open Data has become increasingly important for research purposes. In particular, scholars in the Humanities and cultural heritage institutions leverage the potential of Semantic Web technologies by making cultural heritage data available as Linked Open Data (LOD) \cite{hyvonen_using_2020}. However, data can often be explored via sophisticated search interfaces only \cite{monaco_linked_2022}, which can be overwhelming for users who lack the necessary technical skills.
Data visualisation and storytelling techniques could help experts to share complex information with a broad audience. In the cultural heritage domain, stories help users to navigate large content \cite{fernie_paths_2012}, serve as interpretive frameworks to carry the values and meaning of a culture \cite{bruner_narrative_1991}, and convey complex information compactly \cite{gershon_what_2001, lombardo_storytelling_2012}.

Unfortunately, existing solutions for LOD visualisation are often targeted to experts and present constraints that prevent users with little knowledge of Semantic Web and data engineering to interact with data in creative ways. Several tools focus on schema-level aspects of datasets, and mostly provide interfaces to show one type of visualisation only. Even when tools offer a broad range of visualisations, the narrative element is missing \cite{desimoni_empirical_2020}.

Nonetheless, user-friendly interfaces that allow users to interact with data, create stories, and communicate results effectively, are desirable, in order to increase the awareness and usage of quantitative analysis methods in the humanities and to foster the success of the Open Data business model.

In this paper, we introduce {\itshape Make mE a Linked Open Data storY (MELODY)}, an open-source online platform for querying and presenting charts based on Linked Open Data, combining data-driven results with manually curated content in an article-alike narrative, and publishing web-ready data stories. Editing is supported by {\itshape What You See Is What You Get (WYSIWYG)} interfaces, walk-through, and examples. Different authentication levels allow users to collect stories, to publish them online, and export them in several formats (HTML, PDF, JSON).

MELODY was originally designed to facilitate data dissemination of ten pilot projects dedicated to cultural heritage data. In their project, scholars collect sources (e.g. texts, audio files), extract information, and populate a knowledge graph which can be explored via a number of web interfaces, including both those for lay people and expert enquiries. The methodology used to design MELODY stems on the diversity of such data, the competency questions designed in the requirements elicitation phase, and the tasks of the target audience. Such a methodology is meant to drive the development from ontology design to web design. To accommodate diverse project requirements and to keep a tight dependency with competency questions, MELODY is based on standard Semantic Web technologies and can be reused with any dataset available via SPARQL endpoint.

The remainder of the paper is the following. In section \ref{related_works} we review approaches to design interfaces for interacting with LOD, we identify existing solutions addressing data visualisation, and highlight issues. In section \ref{methodology} we present our hybrid methodology, highlighting its novelty aspects while mapping interaction patterns to competency questions. In section \ref{overview}, we introduce MELODY and its main components. In section \ref{evaluation} we present a competitive analysis to validate requirements of our software. In section \ref{discussion} we discuss results, potential impact, and limitations. We conclude with future works in section \ref{conclusion}.

\section{Related work} \label{related_works}
\subsection{Methodologies for ontology-driven interface design}
In the Semantic Web, the structure of a SPARQL query is closely related to the ontology terms used to organise the data \cite{charalampidis2018semantic}. Queries and users' informative needs drive the design of interfaces for information seeking purposes, and could be mapped to ontology requirements. Therefore, one would expect User Interfaces and Experience (UI/UX) design methodologies to be harmonised with  ontology design practices.

On the one hand, several ontology design methodologies currently adopt tools from Human Computer Interaction (HCI). Such methods adopt a bottom-up approach to requirements elicitation and rely on domain experts when (1) defining the domain space and vocabulary, (2) outlining motivating scenarios, and (3) extracting requirements in the form of natural language Competency Questions (CQs) \cite{gruninger_methodology_1995, noy_state_1997, suarez-figueroa_neon_2015, cota_applications_2020}. The eXtreme Design (XD) methodology \cite{presutti_extreme_2009, carriero_pattern-based_2021} prescribes practises for capturing stakeholder goals, interests, and tasks, and encourage grouping them under high-level categories, i.e. personas \cite{junior_user_2005}, in turn characterised by stories, expectations, and priority levels. Stories are powerful tools for designing experiences as they present facts connected by causal relationships and help frame the motivating aspects of users or to describe unforeseen situations \cite{gruen_use_2002}. So doing, the knowledge acquisition process borrows concepts and tools closer to UI/UX design practices.

On the other hand, it has been demonstrated that ontology-driven approaches have brought significant benefit in reducing interface requirements ambiguity \cite{dermeval_applications_2016, amna_ambiguity_2022} when supporting software development \cite{dermeval_applications_2016, de_souza_improving_2018, yang_ontology-based_2019}, and requirements formulation, e.g. by creating user stories as structured data \cite{thamrongchote_business_2016}. Task ontologies and taxonomies for describing interactive user behaviours and UI elements exist \cite{ramaprasad_design_2014, paulheim_ui_2013, silva_formal_2017}, and have been used to support the assessment of prototypes and final interfaces. Similarly, ontologies and algorithms addressing HCI design in the design of web applications have been discussed \cite{bakaev_application_2016, bakaev_ontology_2010}. While such efforts focus on the description of aspects of the HCI discipline, they do not operationalise the descriptive knowledge of domain ontologies (those used to represent the data and not the UI/UX process) into prescriptive models, i.e. defining how a system is supposed to behave according to ontology requirements (e.g. CQs).

Design Thinking (DT) \cite{rowe_design_1991, brown_design_2008, dorst_core_2011, norman_design_2013} is a well-known design methodology that effectively improves the quality of the ideas generated and reduces the risk of failure \cite{liedtka_evaluating_2017}. It consists of six phases: empathise, define, ideate, prototype, test, and implement. Scholars have tried to encapsulate the logic of DT into ontologies to evaluate design ideas \cite{ramaprasad_design_2014}, to formally describe empathy models \cite{pileggi_knowledge_2021}, or other procedural aspects of HCI \cite{paulheim_ui_2013, silva_formal_2017, bakaev_ontology_2010, bakaev_application_2016}. While such results provide us formal definitions of HCI and DT methods, they do not inform us on how to leverage domain ontologies in the DT process and neglect considerations on the overall, prescriptive framework.

To the best of our knowledge, there is no methodology that supports a research team from the early stages of ontology design to the selection of UI/UX approaches. Nonetheless, this is often the case in projects where knowledge graph generation methods are tied to user-friendly interfaces for information seeking, exploration, and discovery \cite{hyvonen_using_2020}, e.g. projects dealing with the dissemination of cultural heritage on the web \cite{konstantakis_adding_2020}.

\subsection{Linked data visualisation and storytelling}

Presentational aspects of Linked Open Data have attracted the interest of scholarly works in the last decade \cite{brunetti_formal_2013, anutariya_vizlod_2018}. 
In recent studies \cite{desimoni_empirical_2020} 77 data visualisation tools, including those with a focus on Linked Data, have been reviewed. The survey demonstrates that there is a lack of adequate solutions for LOD-based storytelling (YourDataStories\footnote{\url{https://platform.yourdatastories.eu/}} \cite{petasis2018yourdatastories} is the only non-working example), most solutions are not available or focus on schema-level analysis only.

Several problems affect existing solutions, spanning from data access and export, to limitations in charting and storytelling strategies. As highlighted in \cite{bikakis_exploration_2016}, most exploration and visualisation systems work offline (e.g. GraphVizdb \cite{bikakis_graphvizdb_2016}) or are limited to accessing pre-processed, small, datasets (e.g. VisGraph\verb|^|3 \footnote{\url{https://visgraph3.github.io/}}, RDFShape \cite{gayo_rdfshape_2018}). Moreover, access to datasets may be limited to those that are not accessible via SPARQL endpoints or API (e.g. Graphless \cite{santana-perez_graphless_2018}, Rhizomer \cite{brunetti_overview_2013}). Therefore, developing interfaces for the live exploration of large datasets is problematic. When several sources can be explored, tools focus on schema-level analysis (e.g. H-BOLD \cite{po_high-level_2018}, WebVOWL \cite{lohmann_webvowl_2015}) or present RDF data in one format only, often the tabular form (LodView \footnote{\url{https://lodview.it/}}, RDFSurveyor \cite{vega-gorgojo_linked_2019}). Tools that do not face the above issues, still, do not offer the user the possibility to present data in an author-controlled narrative and allow for image download only (e.g. RAWGraphs \footnote{\url{https://www.rawgraphs.io/}}, SynopsViz \cite{bikakis_rdfsynopsviz_2017}). Nonetheless, building a narrative, presenting the data in the right order, and limiting results to selected visualisations that convey the information as simply and concisely as possible, are common methods used in storytelling techniques to bridge users’ cognitive gap \cite{stackpole_next_2020}. Moreover, the What-You-See-Is-What-You-Get (WYSIWYG) paradigm is deemed a powerful way to give the author control while creating such narratives \cite{khalili_rdfa_2012}.

\section{Methodology and requirements collection} \label{methodology}
The diversity of data and audiences of projects dedicated to cultural heritage requires a sound methodological approach to drive the development from knowledge acquisition to web design. We introduce a modular workflow harmonising eXtreme Design and Design Thinking methodologies. The proposed methodology spans from data and user requirements collection to the development of prototypes and their evaluation. In particular, the \textbf{\textit{empathise}} phase of Design Thinking is integrated with methods proposed by eXtreme Design, such as the extraction of competency questions (CQs) from user stories, and the mapping of CQs to ontology design patterns. Secondly, a mapping between CQs and data patterns allows us to gather UI/UX requirements targeted to scholars. Data and interaction patterns guided the design and prototyping of UI components of MELODY, and highlighted alternative strategies for User Experience design.

\subsection{From Ontology Design to Storytelling Interfaces} \label{ontology_to_interface}
\begin{figure}[ht]
    \centering
    \includegraphics[width=0.7\linewidth]{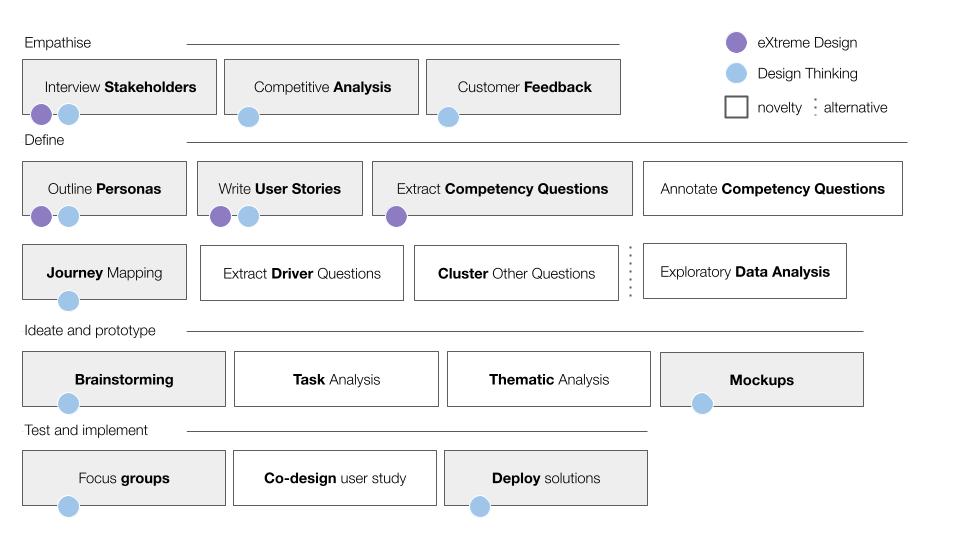}
    \caption{Methodology overview.}
    \label{fig:workflow}
    \Description{A diagram that showcases the phases of the workflow described in \ref{ontology_to_interface}.}
\end{figure}

In Fig. \ref{fig:workflow} we illustrate the phases of our proposed methodology, grouped under the phases of Design Thinking. We highlight methods borrowed from the two methodologies (grey blocks) and our novel contributions (white blocks).

In detail, interviews with stakeholders are performed to define (1) users' motivational aspects, (2) a list of competitors of the product to be developed, (3) user behaviours (journeys), (4) limitations and benefits of competitors, (5) expectations in terms of content and application requirements. Resources mentioned during interviews are included in a competitive analysis, which addresses a classification of aspects relevant to knowledge organisation, User Interface design, and visual identity. A narrow list of reference solutions is further investigated via user studies.

The ontology design team outlines personas, i.e. stereotypical users presented in the form of textual descriptions. Situations, tasks, interests, and expectations, are grouped under one or more user stories for each persona. Competency questions are extracted from stories via content analysis. CQs summarise salient aspects of the story in the form of natural language questions, address ontology and data requirements, and semantic constraints on properties and classes. We propose to further annotate CQs with (1) ontology patterns, including classes and properties of entities involved, (2) a classification of the scope of data (e.g. historical, musical, bibliographical data), (3) whether the CQ addresses a task (e.g. search, share) or secondary detail, and (4) the type of expected result, e.g. strings, lists, tables, or charts.

Stories and CQs are analysed to grasp user journeys, i.e. the flow of searches, reading behaviours, and emotional states. We propose two alternative ways to design journeys from content requirements. If collected personas are rather homogeneous and in a small amount, driver CQs can be identified, and other CQs can be clustered. Driver CQs are those that best summarise the scope of data and the task (e.g. search for artefacts grouped under categories). Other CQs focused on contextual aspects can be grouped (e.g. in info boxes). In case personas, stories, and CQs are too many to be singularly addressed, Exploratory Data Analysis (EDA) can be performed over annotated CQs.

Brainstorming sessions help to make sense of prior results, leading the preparation of mockups. To support this activity, we propose to integrate two analyses: a task analysis, where CQs are aggregated by task, sorted by complexity of the task, and separated accordingly into different interfaces; a thematic analysis, based on the distribution of classes and their position in data patterns (i.e. as input or output), in order to understand common access points to data.

Focus groups with stakeholders involved in the definition of requirements provide meaningful feedback, helping to reframe requirements that were unclear or incorrect. Usability tests and heuristic evaluation with 3-5 people groups are considered sufficient to discover up to 90\% of issues \cite{nielsen1994guerrilla}. To include secondary user targets in the evaluation, we propose a user study focused on co-design. While the initial user study (Customer feedback) guided participants in the evaluation of existing solutions, in a co-design user study, participants have to imagine themselves in a comfortable scenario (e.g. "you are at home studying"), are provided with a task (e.g. "you want to discover new music"), and must share insights on how they would achieve their goal. Results of the survey contribute to a preliminary evaluation of the mockups (despite these are not given to participants). If results are satisfying, mockups are implemented, tested, and solutions are deployed in production.

\subsection{Requirements elicitation}
To design MELODY, we relied on ~220 CQs\footnote{\url{https://github.com/polifonia-project/stories}} extracted from stories of historians, musicologists, and heritage professionals involved in the development of ten pilot datasets, and that later on became early adopters of MELODY. In detail, CQs belong to 19 personas, and were extracted from 28 stories. We annotated the CQs with entities, relations, and potential UI/UX search patterns. Via aggregated data analysis, we analysed the distribution of CQs in terms of entities and properties, and we estimated common data patterns along with recurring interaction patterns (e.g., CQs returning lists, charts, maps). Stories relevant to data visualisation tasks are individually analysed, and driver CQs are identified. We assigned a main UI/UX pattern to each story, including expected visualisations, content types, and tasks\footnote{\url{https://github.com/polifonia-project/web_portal/tree/main/analysis}}. To date, the analysis has focused on three main interaction patterns, namely:
\begin{itemize}
    \item an article-like interface where curated text can be accompanied by charts (e.g. bar charts, pie charts, line charts);
    \item an interactive text search, where the results of a search can feed new searches via reusable queries/components that return new charts (e.g. tables); 
    \item a geographic map, including aids to filter out data points, visualise metadata and statistical information.
\end{itemize}
While such interaction patterns could be used as different, standalone, interfaces, MELODY makes use of all of them into a single template wherein UI components can be combined in a unique narrative.

Functional requirements of the application are also gathered via the competitive analysis, where tools mentioned in section \ref{related_works} and those proposed by stakeholders are analysed according to problems emerged in the literature review. Results of the competitive analysis also guided the evaluation of MELODY.

\section{MELODY: an overview} \label{overview}
{\itshape Make mE a Linked Open Data storY, MELODY} is a web-based, user-friendly platform for (1) querying Linked Open Data available from any SPARQL endpoint, (2) presenting query results as charts, optionally along with curated content, and (3) publishing results as web-ready, exportable, data stories. To create a data story, users need (1) the URL of a SPARQL endpoint and (2) some prior knowledge of SPARQL query language. In fact, any chart and interactive component is derived from an input SPARQL query (written by the user) against the selected endpoint.

In this section we introduce technologies, authentication and authoring mechanisms, and data story components.

MELODY is a light-weight application based on Flask\footnote{\url{https://flask.palletsprojects.com/}} and React\footnote{\url{https://react.dev/}} (non-native) frameworks, that respectively manage backend APIs and templating, and UI components. Any Linked Open Data source available via a public SPARQL endpoint can be queried (currently client-side) and used to build a data story. MELODY can be installed locally or accessed via web browser. The source code, the documentation, and a demo are available online\footnote{Source code: \url{https://github.com/polifonia-project/dashboard}; documentation: \url{https://polifonia-project.github.io/dashboard/}.}.

\begin{figure}[ht]
    \centering
    \includegraphics[width=0.7\linewidth]{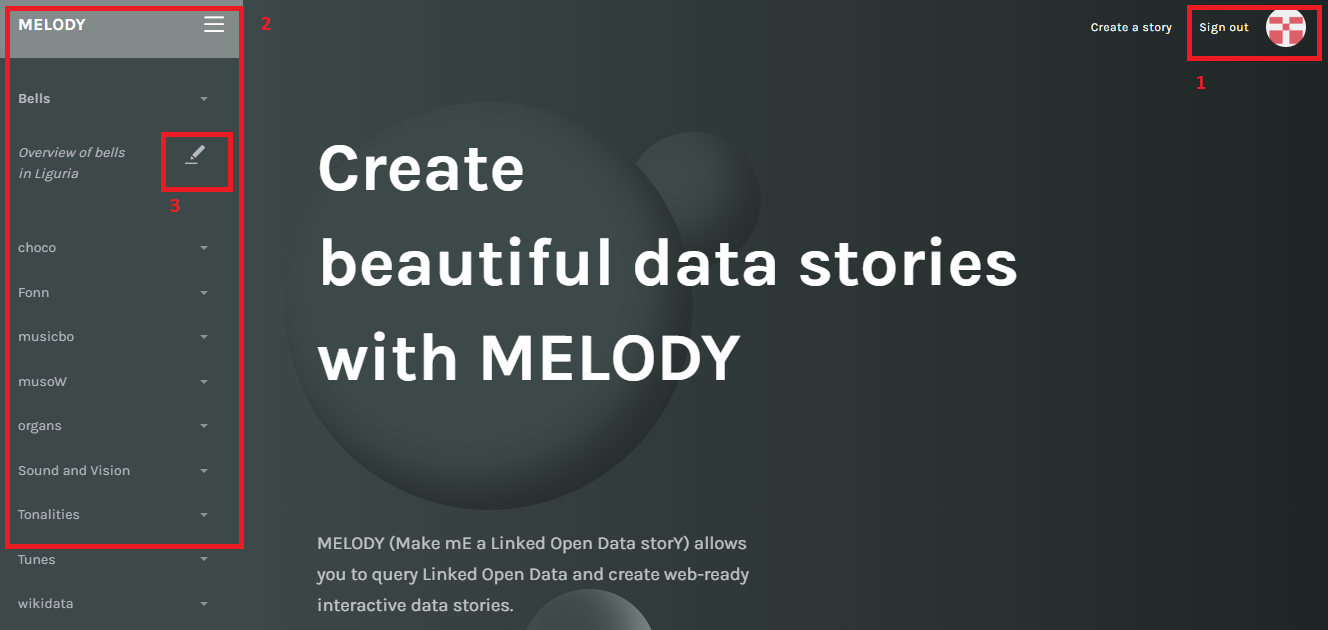}
    \caption{Homepage with authentication (1), sidebar with published data stories (2) and modify button (3).}
    \label{fig:homepage}
    \Description{A screenshot of the Melody homepage that shows the sidebar to access published data stories, the button to modify them, and the login/logout button.}
\end{figure}

Users can create stories anonymously or via a GitHub authentication (Fig. \ref{fig:homepage}), with the only difference being the publication venue of their stories. Any online instance of MELODY must be paired to a GitHub organisation, which includes members of the project maintaining the application, who in turn have access to a self-standing application for authoring and publishing. Data stories created by members of the GitHub organisation are published on the website of MELODY, and are publicly available from the left sidebar. GitHub users that are not organisation members can access any online MELODY instance and use all MELODY features, including the final publication. However, their stories are published as static web documents on a separate catalogue, i.e. a GitHub repository with GitHub pages enabled, that is hosted by the GitHub organisation maintaining MELODY. Indeed, organisation members maintaining the web application may decide to make available their authoring interface to external users, although moving anonymous stories from their instance to a separate, low-maintenance, instance of GitHub pages. For instance, the online demo of MELODY\footnote{\url{https://projects.dharc.unibo.it/melody/}} is paired with an external catalogue, currently used for testing purposes\footnote{\url{https://melody-data.github.io/stories/}}. Finally, anonymous users have access to all the features but the publication of their web-ready data stories, which can only be downloaded at the end of their session.

\begin{figure}[ht]
    \centering
    \includegraphics[width=0.7\linewidth]{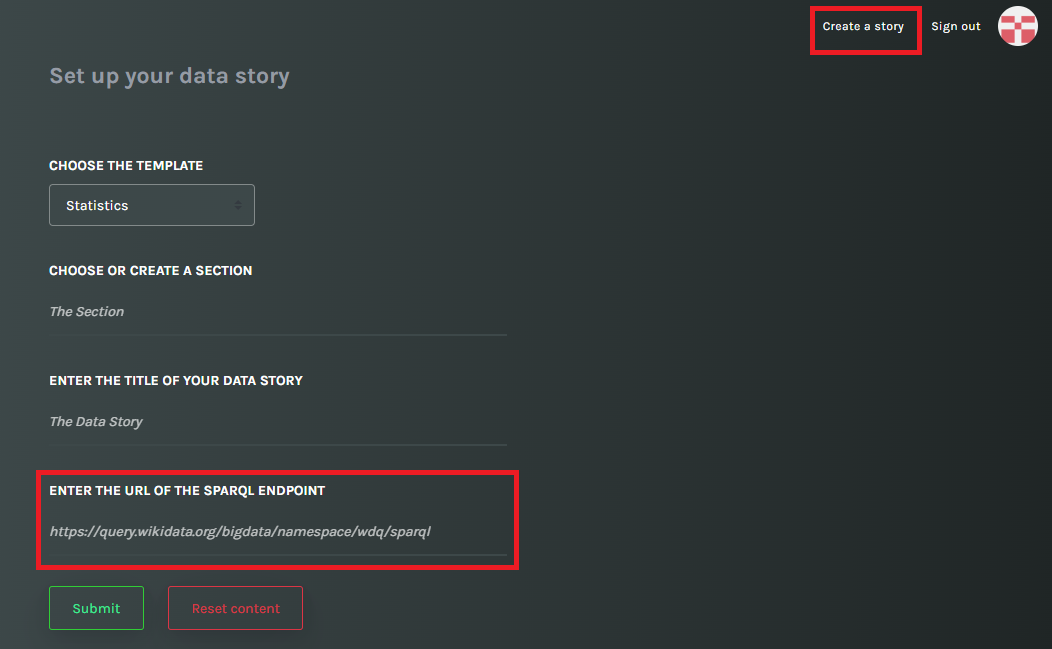}
    \caption{Setup form to start creating a data story.}
    \label{fig:setup}
    \Description{A screenshot that shows a form to be filled in with information about type of template, section in which to include the story, the title and the SPARQL endpoint.}
\end{figure}

As aforementioned, the main requirement to create a data story is the URL of a SPARQL endpoint, and some information to set up the canvas of the data story (Fig. \ref{fig:setup}). Currently, the statistics template includes all the UI components available in MELODY. If the user is a member of the hosting organisation, they can choose or create a section title, which appears in the left sidebar of the website and groups stories under a common theme. An initial title must be provided to initialise the canvas.

\begin{figure}[h]
    \centering
    \includegraphics[width=0.5\linewidth]{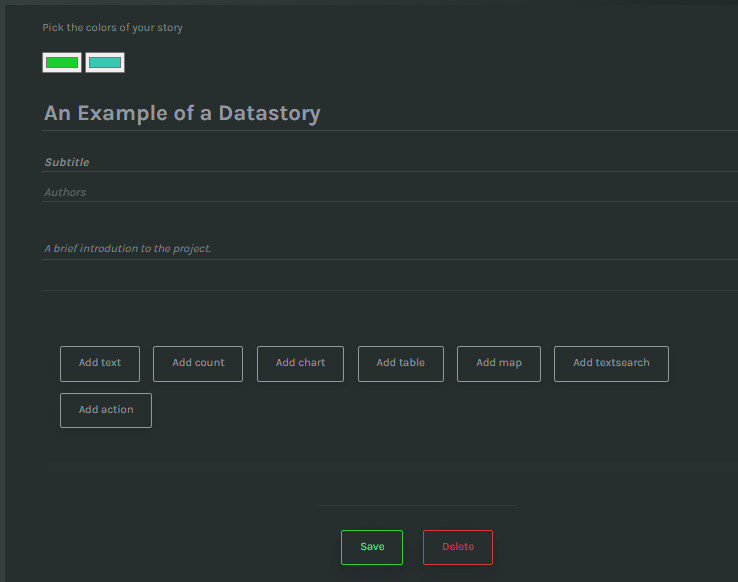}
    \caption{The initial WYSIWYG interface with the list of components.}
    \label{fig:wysiwyg}
    \Description{A screenshot that shows a form to be filled in with information about title, subtitle, description about a story and buttons to add components (text, numbers, charts, table, map, textsearch with actions.}
\end{figure}

\begin{figure}[h]
    \centering
    \includegraphics[width=0.7\linewidth]{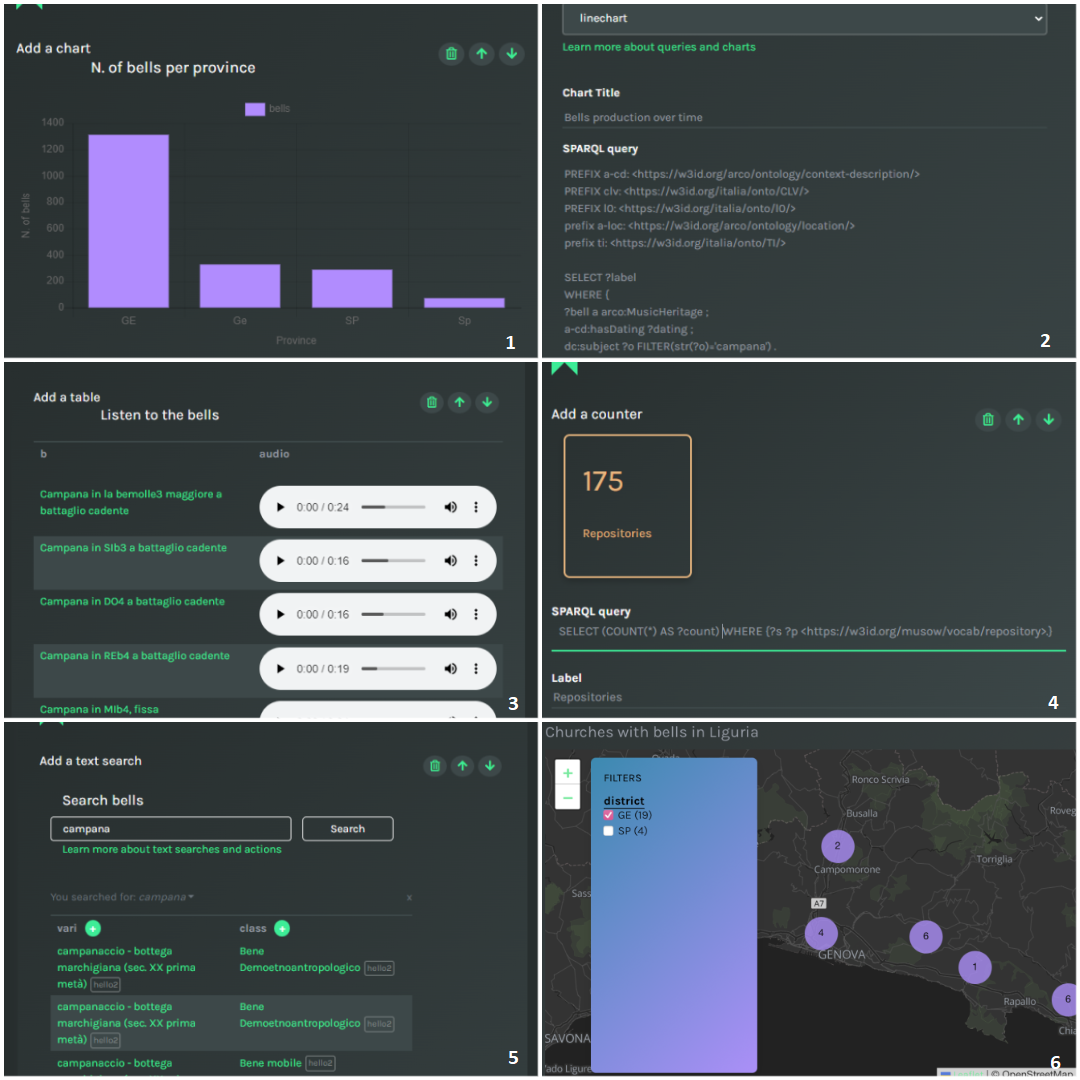}
    \caption{An overview of the components: a chart preview (1), the form for the chart (2), a table preview with audio files (3), a count preview (4), an example of textsearch (5), the preview of the map with filters (6).}
    \label{fig:collage}
    \Description{The image is a collage of all the components in Melody. At the top left a barchart, at the top right a form with type of chart, title and SPARQL query, in the middle left a table with link and audio file entries, in the middle right a form with number, label and SPARQL query, at the bottom left a text search area and a table with link entries, at the bottom right a map centerd in Liguria with data point and a left sidebar that filters by province.}
\end{figure}

The user is redirected to the data story canvas, where they can choose among several components, pick the colour palette of the interactive elements, and start authoring the story by selecting components to be added (Fig. \ref{fig:wysiwyg}). Content creation is simplified by a number of WYSIWYG forms, one for each UI component, that authors can fill in with plain text or SPARQL queries, and instantly see the preview of results (Fig. \ref{fig:collage}). Components respond to requirements extracted during the initial requirements collection. Currently, available components are the following:
\begin{itemize}
    \item text, displaying a rich text editor for creating curated texts, like paragraphs and titles, as HTML content;
    \item counter, displaying a card with a number and a label; 
    \item chart, currently proposing four chart types, i.e. barchart, line chart, scatterplot and doughnut chart; 
    \item map, displaying geolocated data on a map, optionally accompanied by a sidebar with filters and a secondary sidebar to display metadata of data points; 
    \item table, which can include simple text, audio files, or embedded videos; 
    \item text search, displaying a search box for free-text search on a dataset and returning results in a table;
    \item action, a query that can be performed on any result of a text search (i.e. a cell value of the returned table) and that displays new results in another table. Actions can be attached also to results of another action, and can be reused on demand with any value.
\end{itemize}
All components (except the text editor) are accompanied by info boxes that explain fields, naming conventions to be used (if applicable), e.g. variable names, data types of results of a SPARQL query, and examples of usage.

\begin{figure}[h]
    \centering
    \includegraphics[width=0.7\linewidth]{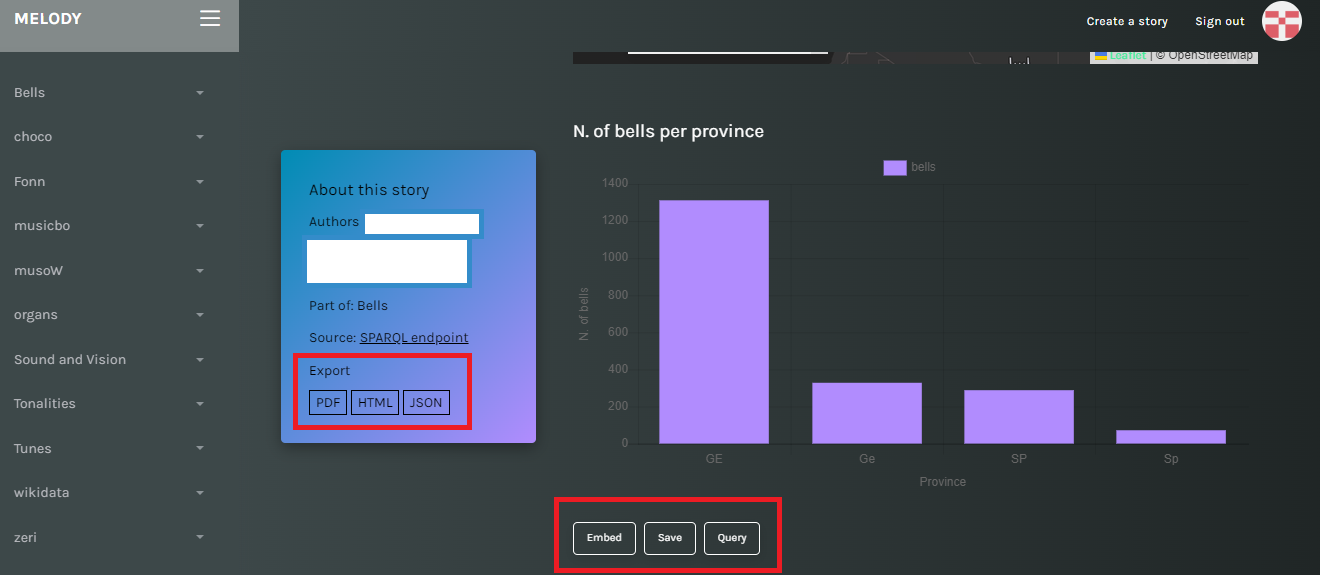}
    \caption{Buttons to export data stories, configuration files, and single components.}
    \label{fig:export}
    \Description{A screenshot that shows a barchart of bells per province, and buttons to export both the chart and the data story as HTML, PDF, JSON, image or embeddable code.}
\end{figure}

The resulting data story is a web document that contains an ordered list of UI components, which can be modified and sorted at any point by authors. All users can export a story as a web-ready static document (HTML), a PDF document, or a JSON file (see Fig. \ref{fig:export}), which keeps track of the order of components in the data story, the SPARQL queries designed by the user, and the type of expected result (e.g. chart, table, HTML text). Likewise, each component of the datastory can be embedded in other web pages, exported as an image or CSV file, and the underlying SPARQL query can be seen on demand.

\section{Evaluation} \label{evaluation}
To validate the soundness of requirements considered to design MELODY, we performed a comparative analysis with competing solutions and collected feedback during focus groups with early adopters.

We selected tools from recent surveys \cite{desimoni_empirical_2020} and we pruned the list by selecting peer-reviewed solutions that (1) support LOD-native interrogation (i.e. SPARQL queries over RDF data), (2) provide at least one charting option as output, (3) are accessible online, (4) are maintained and working, and (5) are openly available for reuse. We ended up with a short list of 9 solutions (Table \ref{tab:results}), that we analysed according to the following aspects:
\begin{itemize}
    \item SPARQL: data can be queried from a SPARQL endpoint;
    \item RDF: data can be queried from a RDF dump;
    \item Multi (datasets): the tool can be reused with any LOD source;
    \item Web: the software provides web-based access;
    \item HTML, Image, embed, CSV, RDF, and PDF: charts or stories can be exported in multiple formats;
    \item Pub: the tool provides a publishing venue of outputs;
    \item Multi (charts): multiple charting solutions are available;
    \item Intrc: interactive charts or query patterns are available (e.g. clickable charts, events triggered by click);
    \item Lay: default pagination of content is provided;
    \item Cur: curated text can be added along with  multiple UI components;
    \item WYSIWYG: facilitation in editing;
    
\end{itemize}

\begin{table}[ht]
\centering
\small
  \caption{A summary of the results of the analysis}
  \label{tab:results}
  \resizebox{\textwidth}{!}{\begin{tabular}{lcccc|ccccccc|cc|ccc}
    \toprule
    \multicolumn{1}{c}{Tool} &
    \multicolumn{4}{c}{Data Access} &
    \multicolumn{7}{c}{Export} &
    \multicolumn{2}{c}{Charting} &
    \multicolumn{3}{c}{Storytelling}\\
    \midrule
    & SPARQL & RDF & multi & Web & HTML & Img & embed & CSV & RDF & PDF & Pub & multi & Intrc & lay & cur & WYSIWYG \\
    \hline
    Graphless & & & & x & & x & & & & & & & x & & &  \\
    \hline
    LDVizWiz & x & & & x & & & & & x & & & x & x & x & &  \\
    \hline
    LodView & x & & x & x & & & & & & & & & & x & &  \\
    \hline
    RAWGraphs & x & x & x & x & & x & & & & & & x & & & & x  \\
    \hline
    RDFShape & x & x & x & x & & x & x & x & & & & & & & &  \\
    \hline
    RDFSurveyor & x & & x & x & & & & & & & & & & x & &  \\
    \hline
    Rhizomer & x & & x & x & & & & & & & x & x & & x & &  \\
    \hline
    Sparklis & x & & x & x & & & & x & & & & & & & &  \\
    \hline
    WebVOWL & & x & x & x & x & x & & & x & & & & x & & &  \\
    \hline
    \textbf{MELODY} & \textbf{x} & & \textbf{x} & \textbf{x} & \textbf{x} & \textbf{x} & \textbf{x} & \textbf{x} & & \textbf{x} & \textbf{x} & \textbf{x} & \textbf{x} & \textbf{x} & \textbf{x} & \textbf{x} \\
    \bottomrule
  \end{tabular}}
\end{table}

In summary, most of the surveyed tools provide a web interface to access multiple data sources via their SPARQL endpoint. Fewer solutions also allow access to RDF dumps, usually as an additional option.
Outputs can usually be exported as images or CSV, in a few cases as RDF data. Little support is offered when downloading results in other formats, such as HTML, embeddable components, or PDF documents. No solution is designed to host or publish results on the web, which is delegated to the user and external venues (LodView is designed as a web application, which can serve data, but it requires the user to install and maintain both the application and results).
Moreover, only a third of tools offer more than one output (interactive) chart. 4 tools out of 9 allow the automatic pagination of content in a canvas, though the control is entirely on the interface side. Notably, none of the surveyed solutions allow curated content (other than titles and legend) to be authored and displayed along with the charts, and only one solution (RAWGraphs) facilitates chart tuning via WYSIWYG interfaces.

\section{Discussion} \label{discussion}
The comparative analysis shows that MELODY is a competitive solution for Linked Open Data visualisation and storytelling. Although MELODY was initially developed to support requirements of pilot datasets that are all part of the cultural heritage domain, the platform can be used with any LOD source available via a SPARQL endpoint, and provides generic interaction patterns to accommodate needs of diverse projects. It is flexible to several case studies, since it facilitates authoring with user-friendly interfaces, allows curated content to be included in an article-alike web document, offers several options for charting and exporting outputs, and acts as a venue for web publication.

It is worth noting that storytelling and authoring features implemented in MELODY can lower the learning curve that is usually required by visualisation tools, since it only requires prior knowledge of SPARQL query language. Despite the reputation of SPARQL of being a language for tech-savvy people only, recent studies on knowledge transfer in interdisciplinary projects demonstrate that it is by far the most suitable solution to empower humanists that want to face complex, data-driven, enquiries, which are rarely addressed by general purpose portals \cite{burrows2022medieval}. While MELODY is not targeted to lay users, it can support beginner-intermediate scholars learning Semantic Web technologies.

The potential impact of MELODY is also demonstrated by the possibility to access for free any online instance of the tool that makes available a publication venue for external users via GitHub pages. So doing, authors can rely on existing, low cost publishing solutions. Moreover, results are made available as static HTML pages with no technology dependency, which makes it easier to move, modify, and maintain the code for users with basic web development skills (i.e. HTML and CSS skills). Such peculiarities could be leveraged in scholarly article authoring environments, although some compelling requirements are currently missing (e.g. citability and dereferencing of new identifiers).

A few limitations affect the current work. Compared to surveyed tools, neither access nor export of rdf static files is currently possible. To date, only SPARQL queries to one endpoint at time can be performed. Federated queries can be leveraged to perform some reconciliation and graph construction on the fly. However, queries are performed client-side and query results are not stored. Therefore, the application may face performance issues when querying large datasets. Moreover, live queries imply modifiability of results, therefore hampering the citability of the data story itself. Lastly, focus groups with early adopters of MELODY provided feedback for validating the online instance, which currently includes the three aforementioned interaction patterns (article-alike, geographic exploration and interactive text search). So far, around ten data stories have been produced by members of pilot projects, and feedback has been collected in GitHub issues to guide the development and enhancement. Follow-up workshops with new, external, participants are planned to estimate the easiness and flexibility of the tool within groups with different technological skills (e.g. graduated students in Digital Humanities).

\section{Conclusions} \label{conclusion}
In this article we presented MELODY, a platform for Linked Open Data storytelling. The design of the application was guided by a novel, hybrid, ontology-driven methodology which leverages requirements expressed in the form of competency questions. The set of competency questions selected is relevant to ten pilot projects in the cultural heritage domain, ensuring a degree of flexibility of realised components. Compared to existing software solutions, MELODY outperforms in most of the functional requirements we collected from the literature and from interviews with stakeholders. 
In future works we will address current limitations related to data access performance, citability of data stories, and we will extend the methodology with new pilot projects to collect new requirements and enrich the platform with new interaction patterns.

\begin{acks}
This work has received funding from the European Union’s Horizon 2020 research and innovation programme (Polifonia, G.A. 101004746).
\end{acks}

\bibliographystyle{ACM-Reference-Format}
\bibliography{main}

\end{document}